\documentclass[11pt,twoside]{article}

\usepackage{asp2006}
\usepackage{epsf}
\usepackage{psfig}
\usepackage{lscape}

\begin{document}
\title{The class of interacting binaries Double Periodic Variables}   
\author{Ronald Mennickent\altaffilmark{1} \& Zbigniew Ko{\l}aczkowski\altaffilmark{1,2} }   
\altaffiltext{1}{Departamento de Astronom\'{\i}a, Universidad de Concepci\'on, Casilla 160-C, Concepci\'on, Chile}
\altaffiltext{2}{Instytut Astronomiczny Uniwersytetu Wroclawskiego, Kopernika 11, 51-622 Wroclaw, Poland
}

\begin{abstract} 
We introduce the class of intermediate mass binaries named Double Periodic Variables (DPVs), characterized by orbital photometric variability (ellipsoidal or eclipsing) in time scales of few days and a long photometric cycle 
lasting roughly 33 times the orbital period. After a search conducted in the OGLE and ASAS catalogues, we identified 114 of these systems in the Magellanic Clouds and 11 in the Galaxy. We present results of our photometric and spectroscopic campaigns on DPVs conducted during the last years, outlining their main observational characteristics. We present convincing evidence supporting the view that DPVs are semidetached interacting binaries with optically thick discs around the gainer, that experience regular cycles of mass loss into the interstellar medium. The mechanism regulating this long-term process still is unknown but probably is related to relaxation cycles of the circumprimary disc.  A key observational fact is the modulation of the $FWHM$ of He\,I 5875 with the long cycle in V\,393\,Sco. The DPV evolution stage is investigated along with their relationship to Algols and W Serpentid stars. We conclude that DPVs can be used to test models of non-conservative binary evolution including the formation of circumbinary discs. 
\end{abstract}


\section{History and prehistory of DPVs}   

DPVs were discovered in the Small Magellanic Cloud after a search for Be stars in the OGLE-II database (Mennickent et al. 2003). They were clearly distinguished from other variables
by showing 2 linked photometric cycles ($P_{o}$ and $P_{l}$). A spectroscopic monitoring of some of them allowed to associate the short periodicity to the orbital period of a binary (Mennickent et al. 2005) whereas 
the long term variations were found to be reddish and non strictly constant (Mennickent, Assman, \& Sabogal 2006, Michalska et al.\,2009).   

The current census of DPVs amounts to 114 in the Magellanic Clouds and 11 in our Galaxy (see magnitud-color and period-period diagrams in Mennickent \& Ko{\l}aczkowski 2009a).  After the discovery of additional variability in V\,393\,Sco (Pilecki 
\& Szczygiel 2007), we recognized it as the first DPV in our Galaxy. Later we found that in the past an additional long cycle was also reported for the galactic DPV AU\,Mon (Lorenzi 1985). Both stars were also found during our independent search for Galactic DPVs in the ASAS database.

\begin{figure}[!ht]
\plotone{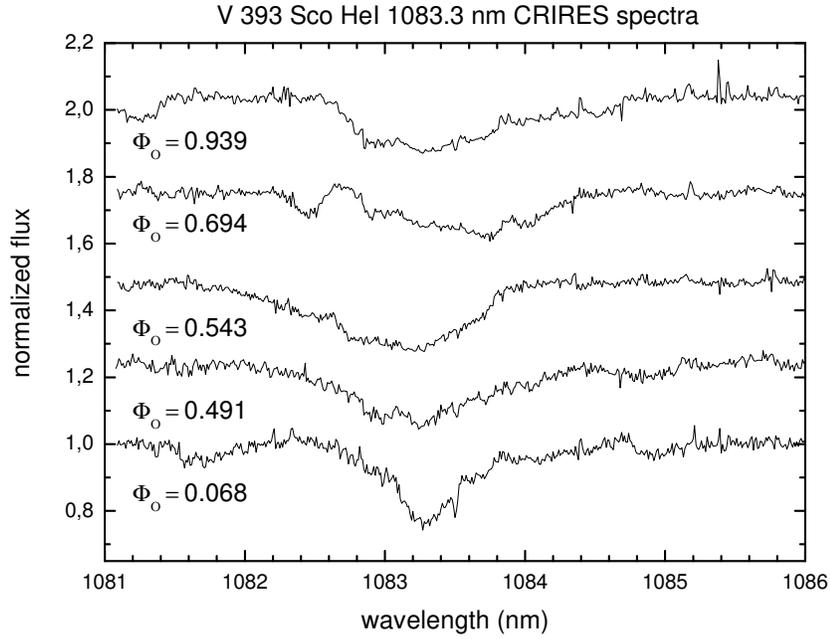}
\caption{Infrared spectra of V\,393\,Sco showing signatures of outflow (blue wing depression around phase 0.54) and gas stream (red wing depression around phase 0.94). The emission feature at the blue wing around phase 0.69 appears in several spectra and could be associated to a bright spot. }
\end{figure}

\begin{figure}[!ht]
\plottwo{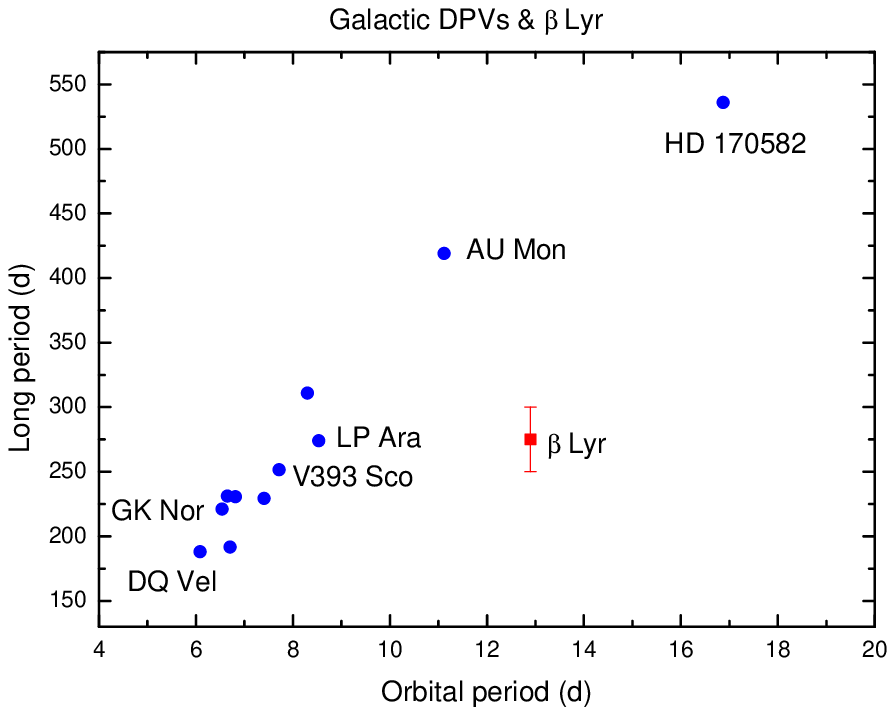}{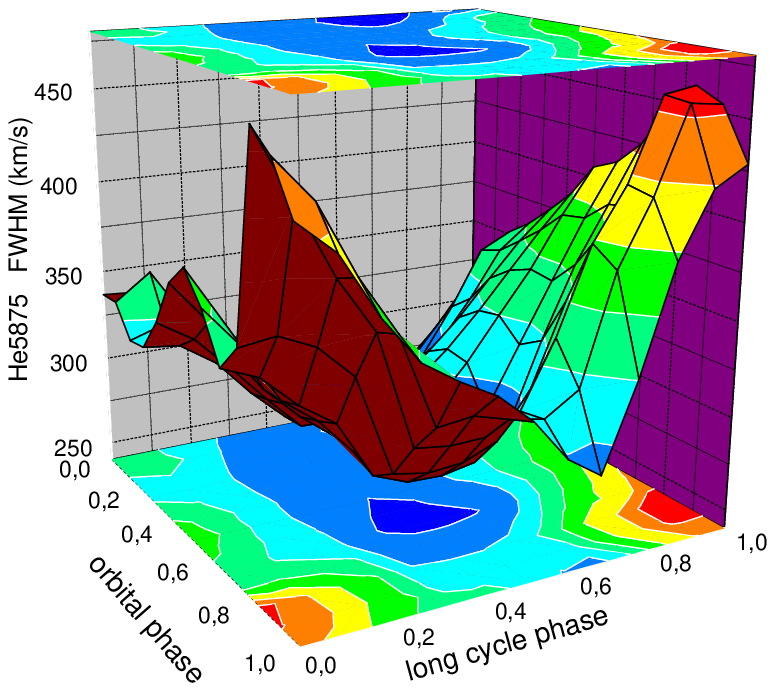}
\caption{Left: Galactic DPVs along with $\beta$ Lyrae. Right: The strong modulation of the $FWHM$ for the He\,5875 line in V\,393\,Sco during long cycle probably  reflects a rotational instability of the circumprimary disc. }
\end{figure}

\section{Evidence for mass exchange and mass loss in DPVs}   

Few DPVs have been  studied in detail, but we can get insights on DPVs as a class based on well studied and representative cases. 
Cumulative evidence indicates that DPVs are interacting binaries with a component (the donor) filling their Roche lobe and transferring mass to the gainer (the primary).
Broad and variable He\,I lines probably probe an accretion disc that sometimes hides the primary.
H\,I  (sometimes He\,I) line emission is the rule (although usually not quite prominent).
It is probable that the deeper DPV eclipse corresponds mostly to the occultation of  the circumprimary disc. 
We observed a loop in the color-magnitude diagram of LMC-DPV1 
during the long cycle that  interpreted in terms of mass loss (Mennickent et al. 2008, hereafter M08). The same star shows discrete Pa$\beta$ and Pa$\gamma$  absorption components following a saw-teeth
pattern with the orbital period indicating outflows through the outer Lagrangian points (M08). The same phenomenon could explain the depressed blue wings observed in the He\,I 10833 \AA~ infrared spectra of V\,393\,Sco near secondary eclipse (Fig.\,1). We observe (minor) variability in the shape of the light curve for V\,393\,Sco during main minima that could indicate changes in the properties of the circumprimary disc. In addition, the H$\alpha$ emission line strength increases during supercycle maximum (Mennickent \& Ko{\l}aczkowski 2009b). Most DPVs with 2MASS data seems to show infrared excess. In the studied cases  mass ratios (donor/gainer) are always less than one. All these singular characteristics, plus the presence of two distinct periodicities, suggest that DPVs can be observed as a new class of interacting binaries, at least from the observational point of view.

\section{The hidden instability in DPVs}

We propose that DPVs are Case-A/B mass transfer binaries after mass ratio reversal in Algol-like configurations. They are more massive than ordinary Algols (Mennickent \& Ko{\l}aczkowski 2009a), so it is possible that
the mass transfer rate is larger, and the primary is rotating at critical velocity. Under these circumstances, accretion is stopped and the disc starts cumulating mass that is periodically ejected from the system. Our observations indicate that mass loss occurs permanently in DPVs, mainly through the outer Lagrangian points. However, the long term periodicity implies that there is another clock governing mass loss in the long term. We believe that during the supercycle the disc cumulates extra matter that is is expelled from the binary during supermaximum. The remarkable behavior of He\,I in V\,393\,Sco (Fig.\,2) suggests  that the rotational velocity of the circumprimary disc is larger during supermaximum and modulated with supercycle phase. 
The mechanism for this supercycle is unknown. We analyzed the possibility that the disc outer radius grows until the 3:1 resonance radius and disc starts to precess, as happens in low mass ratio SU\,UMa stars (Mennickent \& Ko{\l}aczkowski 2009a). In this view 
precession enhances mass loss into the interstellar medium. However, the fact that $FWHM$ maintains the same orbital behavior during supercycle (Fig.\,2)  suggests that there is no disc precession. Other more speculative hypothesis is that the primary experiences instabilities around critical velocity, gaining extra momentum during accretion until attains a velocity just above the critical one, then relaxes below critical velocity giving the extra momentum to the disc that partly escapes from the binary. Additional studies are needed to confirm this view. We have initiated a program to study DPVs with high resolution optical/infrared spectrographs and robotic telescopes with the aim of shedding light on this phenomenon.

\section{Relationship to Algols and W Serpentids}

In Table 1 we summarize our view  for DPVs in the context of Algols and W Serpentid stars. It is possible that the critical velocity of the gainer can be maintained only until the mass
of the donor ($M_{2}$) drops to certain value. During this period of high mass transfer rate  the star behaves as a W Serpentid (with a thick circumprimary disc and chaotic mass loss) or as a DPV (with slightly lower $\dot{M_{2}}$  allowing the DPV instability to operate). When $\dot{M_{2}}$ drops even more,  tidal forces spin down the gainer, the orbital separation ($a$) increases, and the system becomes a typical Algol star. In Algols mass transfer rate if present is comparatively small, partly due to the less massive donor and also to the larger $a$. 
$\beta$ Lyr probably still is not a DPV, as suggested by  their very small long period amplitude and position in the $P_{l}-P_{o}$ diagram (Fig.\,1).  



\begin{table}[!ht]
\caption{Proposed observational/evolutionary link between W Serpentids, DPVs and Algols.  The critical rotational velocity of the gainer is labeled $v_{c}$.}
\smallskip
\begin{center}
{\small
\begin{tabular}{ccccc}
\tableline
\noalign{\smallskip}
Systems & Key facts & $M_{2}$, $\dot{M_{2}}$ & Mass loss & Age, $v_{c}$ \\
\noalign{\smallskip}
\tableline
\noalign{\smallskip}
W Ser & polar jets, variable eclipse, large  $\dot{P_{o}}$  & large & large& young, yes\\
DPV & 2-periods, small ecl. variability, $\dot{P_{o}} \approx 0$   & medium & cyclic& middle, yes\\
Algol &small/no additional variability, $\dot{P_{o}} \approx 0$  & small &small& old, no\\
\noalign{\smallskip}
\tableline
\end{tabular}
}
\end{center}
\end{table}

\acknowledgements REM acknowledges support by Fondecyt grant 1070705, the Chilean 
Center for Astrophysics FONDAP 15010003 and  from the BASAL
Centro de Astrof\'isica y Tecnologias Afines (CATA) PFB--06/2007.


\noindent
{\it S.M. Rucinski}: What about blends? They are always of concern in other ga\-laxies. \\
\noindent
{\it R.E. Mennickent}: DPVs are also observed in the Galaxy. We have discarted  the possibility of blends in the case of DPVs.  \\

\end{document}